\documentclass{article}
\usepackage{arxiv}

\usepackage{geometry}
\usepackage{graphicx}
\usepackage[hidelinks]{hyperref}

\usepackage{amssymb}
\usepackage{bm}
\usepackage{lineno}
\usepackage{amsmath}

\usepackage{multirow}
\usepackage{float}
\usepackage{tabularx}
\usepackage{subcaption}
\usepackage{booktabs}
\usepackage{array,xcolor}
\usepackage{arydshln}

\usepackage{algorithm, algpseudocode}

\usepackage{cite}
\title{NeuroPINNs: Neuroscience Inspired Physics Informed Neural Networks}
\author{
  Shailesh Garg  \\
  Department of Applied Mechanics\\
  Indian Institute of Technology Delhi\\
  Hauz Khas, New Delhi 110016, India. \\
  \texttt{shailesh.garg@am.iitd.ac.in} \\
  \And
  Souvik Chakraborty  \\
  Department of Applied Mechanics\\
  Yardi School of Artificial Intelligence (ScAI)\\
  Indian Institute of Technology Delhi\\
  Hauz Khas, New Delhi 110016, India. \\
  \texttt{souvik@am.iitd.ac.in}}
\begin{document}
\maketitle
\begin{abstract}
We introduce NeuroPINNs, a neuroscience-inspired extension of Physics-Informed Neural Networks (PINNs) that incorporates biologically motivated spiking neuron models to achieve energy-efficient PDE solving. Unlike conventional PINNs, which rely on continuously firing activations and therefore incur high computational and energy costs, NeuroPINNs leverage Variable Spiking Neurons (VSNs) to enable sparse, event-driven communication. This makes them particularly well-suited for deployment on neuromorphic hardware and for scenarios with constrained computational resources, such as embedded and edge devices. A central challenge, however, lies in reconciling the discontinuous dynamics of spiking neurons with the smooth residual-based loss formulation required in PINNs. Direct smoothing introduces systematic biases, leading to inaccurate PDE learning. To overcome this, we employ a novel stochastic projection method inspired from upscaled theory that faithfully captures spiking behavior while maintaining compatibility with gradient-based optimization. Standard surrogate backpropagation is used for parameter updates, ensuring computational tractability. We demonstrate the effectiveness of NeuroPINNs on four representative PDE problems across both regular and irregular domains. Furthermore, application of NeuroPINN for linear elastic micromechnics in three dimensions was also explored. 
Results show that NeuroPINNs achieve high accuracy while substantially reducing communication and energy demands, marking a step toward scalable, neuromorphic-ready scientific machine learning.
\end{abstract}
\section{Introduction}
Partial Differential Equations (PDEs) \cite{evans2022partial,strauss2007partial} are employed in various fields of science and technology to model the physical space around us. Their ability as a tool to account for spatial and temporal variations, and, boundary and initial conditions, makes them particularly valuable for this purpose. However, this flexibility also increases the complexity of the equation, and as the underlying process being mapped becomes increasingly complex, solving PDEs becomes an intricate task. In the literature, only a small set of PDEs boast an analytical solution, and often, numerical methods are used to solve such equations.
Among various numerical methods, finite element \cite{larson2013finite,johnson2009numerical}, finite difference \cite{thomas2013numerical,strikwerda2004finite}, finite volume \cite{moukalled2016finite,mazumder2015numerical}, and, spectral methods \cite{bueno2006spectral,kopriva2009implementing} are few that have emerged as gold standards and that can address a wide variety of PDEs with high accuracy. Individual methods offer distinct advantages, and the choice of one over the other is governed by the type of problem at hand.

While accurate, these traditional numerical techniques suffer from high computational cost, which makes them impractical for solving complex PDEs efficiently.
A viable alternative to these traditional methods is the Physics Informed Neural Networks \cite{raissi2019physics} (PINNs). While the underlying basics for PINNs have existed in the literature \cite{dissanayake1994neural,lagaris1998artificial}, their contemporary version, proposed in \cite{raissi2019physics}, has gained significant traction in various fields \cite{cai2021physics,huang2022applications,liu2020generic,cai2021physics_review} of science and technology. PINNs represent the solution to the PDE as a neural network and use its gradients to compute the PDE's residual, to be used as their loss function for training. Because PINNs are composed of basic deep learning \cite{lecun2015deep,goodfellow2016deep} components like feed-forward layers and continuous activations, they suffer from the general drawbacks of deep learning like high energy budget, long training times, and sensitivity to hyperparameters. In particular, we address the issue of excessive energy consumption of PINN architectures because of their use of continuous activations within artificial neurons. 

The high energy demands of Physics-Informed Neural Networks (PINNs), and deep learning models more broadly, limit their scalability and restrict their use in edge computing applications. Motivated by the remarkable energy efficiency of neuroscience-inspired spiking neuron models, we propose NeuroPINNs: a new class of PINNs that integrate Variable Spiking Neurons (VSNs) into their architecture. VSNs emulate biological neurons by accumulating incoming signals and transmitting them only when a threshold is reached, thereby enabling sparse, event-driven communication. Unlike conventional spiking models, however, VSNs propagate information through continuous activations once triggered, avoiding the rigidity of binary spikes while preserving efficiency. By replacing standard continuous activations with VSNs, NeuroPINNs significantly reduce energy consumption while retaining expressive power. Yet, incorporating VSNs introduces two central challenges for training, which we address in this work.

The first challenge stems from the discontinuous dynamics of VSNs, which make direct gradient computation during backpropagation infeasible. To address this, we employ surrogate gradient learning, where discontinuities are approximated by smooth surrogate functions whose gradients can be backpropagated efficiently. This enables training while retaining the underlying spiking behavior. Existing approaches in the literature often rely on conversion strategies, where conventional artificial neural networks are trained first and then transformed into spiking counterparts. However, such methods typically demand long spike trains to achieve competitive performance, which undermines their energy efficiency. Moreover, not all continuous activations can be reliably converted to spiking neurons without imposing strong assumptions or approximations. In contrast, natively trained spiking neural networks (as in our approach) naturally integrate spiking dynamics during learning and consistently yield better performance, while fully exploiting the benefits of event-driven communication.
The second challenge concerns the computation of the residual-based loss function, which is central to the training of PINNs. Unlike conventional neural networks that learn directly from data, PINNs enforce the governing PDEs by minimizing the residuals, which requires differentiating the network output with respect to its inputs. With VSNs, however, the discontinuities in their dynamics prevent direct use of gradients for residual evaluation. While surrogate gradients could, in principle, be applied, doing so introduces approximation errors into the loss and risks steering the network toward solving a distorted version of the PDE. To overcome this, we adopt an upscale theory-inspired Stochastic Projection (SP) method, which has been shown to perform effectively within PINN architectures. SP enables the faithful computation of gradients needed for the residual loss, thereby allowing NeuroPINNs to be trained natively without corrupting the underlying physics. Importantly, surrogate backpropagation remains necessary within NeuroPINNs, but only for parameter updates, while the SP method ensures that the PDE residuals are computed accurately.

In the existing literature, two other spiking PINN variants have been proposed. \cite{tandale2024physics} introduced a physics-based self-learning hybrid spiking neural network for nonlinear viscoplastic FEM simulations, integrating Leaky Integrate-and-Fire (LIF) and recurrent LIF neurons within PINN architectures. However, their approach does not address the critical challenge of computing the residual-based loss. Furthermore, vanilla spiking neurons such as LIF have generally performed better in classification tasks than in regression, which limits their suitability for PDE learning. For this reason, NeuroPINNs employ Variable Spiking Neurons (VSNs) \cite{garg2023part1,garg2024neuroscience,jain2025hybrid,garg2025distribution}, which generate nonbinary spikes and have consistently demonstrated superior performance over LIF neurons in regression tasks. Another variant, proposed by  \cite{theilman2024spiking}, utilizes modified integrate-and-fire neurons but relies on a conversion strategy, where pretrained vanilla PINNs are transformed into their spiking counterparts. This conversion process requires assumptions and approximations, which can introduce errors in the final predictions. For clarity, we refer to this class of models as Converted PINNs (CPINNs) in the remainder of this work. The highlights of the proposed NeuroPINNs can be summarized as follows,
\begin{itemize}
    \item \textbf{VSNs as spiking neurons}: The proposed NeuroPINNs use VSNs within their architecture that are posed to perform well in regression tasks compared to the popularly used Leaky Integrate and Fire (LIF) neurons used in literature for SNNs. Through the use of graded spikes, VSNs can take advantage of both sparse communication and continuous activations. Authors would like to note here that the use of graded spikes can reduce energy efficiency compared to when using vanilla LIF neurons; however, energy efficiency is hypothesized to be more than when using vanilla artificial neurons. 
    \item \textbf{SP method for computing the PINN's residual loss}: When computing the loss function in PINN architectures, gradients of the network representing the PDE solution are required. Using methods like surrogate backpropagation, which change the makeup of the network during backpropagation, can hence introduce inconsistencies during training. SP method does not employ backpropagation and retains the integrity of the network when computing the gradients to be used in the PINN's loss function. This makes it suitable for use in spiking PINN architectures as the network's behaviour remains unchanged while computing gradients.
    \item \textbf{Natively trained in spiking domain}: The proposed NeuroPINNs are trained natively. SP method is used to compute the gradients to be used in the residual loss function, and surrogate backpropagation is used to compute the gradients of the residual loss function with respect to the trainable parameters during training. Natively training any SNN in general is shown to require smaller spike trains compared to converted SNNs, and this also avoids the assumptions taken during the conversion process.
\end{itemize}
We discuss four different PDE examples, along with an application to a three-dimensional linear elastic micromechanics problem, to test the efficacy of the proposed framework. The results produced advocate favorably for the performance of the proposed framework. We compare the performance of NeuroPINNs against the performance of vanilla SP-PINNs and against NeuroPINNs that utilize surrogate backpropagation to compute the loss function. These are referred to as Surrogate Backpropagation NeuroPINNs (SB-NeuroPINNs) in the following text. We have also drawn a comparison against CPINNs, the variant of Spiking PINNs, proposed in \cite{theilman2024spiking}.

The remainder of the paper is organized as follows. Section \ref{section: Proposed Approach} presents the proposed NeuroPINN framework in detail, including the integration of Variable Spiking Neurons (VSNs), the challenges arising from their discontinuous dynamics, and the strategies we employ to overcome them. Section \ref{section: numerical} showcases a series of numerical experiments on representative PDEs defined over both regular and irregular domains, highlighting the accuracy, efficiency, and scalability of NeuroPINNs in comparison to existing approaches. Finally, Section \ref{section: Conclusion} summarizes the key findings, reflects on the broader implications of this work for scientific machine learning on neuromorphic hardware, and outlines potential avenues for future research.

\section{Neuroscience Inspired Physics Informed Neural Network}\label{section: Proposed Approach}
In this section, we present the details of the proposed Neuroscience-Inspired Physics-Informed Neural Networks (NeuroPINNs) for solving partial differential equations (PDEs). Similar to the original PINN framework introduced in \cite{raissi2019physics}, NeuroPINNs are deep learning architectures that learn the solution of a PDE by enforcing the underlying physical laws rather than relying solely on data. The distinguishing feature of NeuroPINNs lies in their use of neuroscience-inspired neuron models within the network architecture, which substantially reduce the energy requirements during training and inference, and can potentially enable efficient deployment on event-driven hardware such as neuromorphic chips. As in standard PINNs, the learning process is guided by a physics-based loss function derived from the residual form of the governing PDEs, ensuring that the network predictions remain consistent with the physical system being modeled. Consider a general form of a governing partial differential equation,
\begin{equation}
\begin{gathered}
    \mathcal{N}\left(\bm x, t, u, \partial_{\mathrm{t}} u, \partial_{\mathrm{t}}^2 u \ldots, \partial_x u, \partial_{\mathrm{t}}^n u, \ldots, \partial_x^n u; \bm\alpha_\mathcal N\right)=0, \bm x \in \Omega, t\in (0,T],\\
    u(\bm x,0) = f_i(\bm x), \bm x \in \Omega,\\
    \mathcal B(u(\bm x, t); \alpha_\mathcal B) = f_b(\bm x, t), \bm x \in \partial\Omega, t\in (0,T],
\end{gathered}
\end{equation}
where $u$ denotes the field variable of interest, defined on the spatial domain $\Omega$ with boundary $\partial\Omega$, and evolved over the temporal interval $\left(0,T\right]$. The operator $\mathcal N$ represents the nonlinear differential operator governing the dynamics of the system, parameterized by $\bm \alpha_\mathcal N$, and involves derivatives of $u$ with respect to both spatial and temporal coordinates. Similarly, $\mathcal B$ denotes the boundary operator, parameterized by $\bm \alpha_\mathcal B$, which encodes the boundary conditions of the problem. The functions $f_i(\bm x)$ and $f_b(\bm x, t)$ specify the prescribed initial and boundary conditions, respectively. 

A NeuroPINN architecture is constructed using standard deep learning layers (e.g., dense layers) to approximate the field variable $u$ of the governing equation. The network is denoted as $\mathcal{U}(\bm x, t; \bm w)$, where $\bm{w}$ represents the trainable parameters, such that
\begin{equation}
    u\simeq\mathcal U(\bm x, t; \bm w),
\end{equation}
To enhance energy efficiency, the NeuroPINN replaces conventional continuous activation functions with Variable Spiking Neurons (VSNs). VSNs promote sparse, event-driven communication, thereby reducing computational load and energy requirements while potentially enabling deployment on neuromorphic hardware. The dynamics of a VSN are defined as, 
\begin{equation}
\begin{gathered}    
    \mathcal M_{\bar t} = \beta_l \mathcal M_{\bar t-1} + z_{\bar t},\\
    \widetilde y = \left\{ \begin{matrix}1, & \mathcal M_{\bar t}\geq T_h\\
    0, & \mathcal M_{\bar t}<T_h\end{matrix}\right.,\\
    \text{if } \widetilde y = 1, \mathcal M_{\bar t} \leftarrow 0,\\
    y_{\bar t} = \sigma(\widetilde y\,z_{\bar t}), \text{ given } \sigma(0) = 0,
\end{gathered}
\end{equation}
where $y_{\bar t}$ is the output of the VSN corresponding to the input $z_{\bar t}$ at the ${\bar t}$\textsuperscript{th} Spike Time Step (STS) of the input spike train\footnote{Information in spiking neurons is transmitted as spike trains, which are sequences of spiking and non-spiking events over time. Together, these events encode a real-valued quantity, with the discrete index $\bar t$ denoting the time step at which each event occurs. This time step may also be referred to as the Spike Time Step (STS).}. The memory variable $\mathcal M_{\bar t}$ represents the accumulated potential of the VSN at the  ${\bar t}$\textsuperscript{th} STS, , which resets to zero (or undergoes a soft reset by reducing
$\mathcal M_{\bar t}$ by a fraction) whenever the threshold $T_h$ is crossed. The parameters $\beta_l$ (leakage factor) and $T_h$ may be treated as trainable during learning. In practice, VSNs replace continuous activations in the earlier layers of NeuroPINNs to enforce event-driven computation, while the final layers retain conventional activations to stabilize regression accuracy.

Training a NeuroPINN requires defining a loss function motivated by the underlying PDE. Specifically, the total loss is expressed as a weighted combination of physics-informed, boundary, and initial condition residuals, 
\begin{equation}
\begin{gathered}
    \mathcal{L}_{P D E}=\left\|\mathcal{N}(\mathcal U(;\bm w); \alpha_\mathcal N)\right\|_{\Omega\times(0, T]}^2, \\
    \mathcal{L}_{B C}=\| \mathcal{B}\left(\mathcal { U }
    (\cdot;\bm w); \alpha_\mathcal B)-f_b(\cdot) \|_{\partial \Omega \times(0, T]}^2\right., \\
    \mathcal{L}_{I C}=\left\|\mathcal U(\cdot
    ;\bm w)-f_i(\cdot)\right\|_{\Omega}^2, \\
    \mathcal{L}_{\text {Total }}=\lambda_{P D E}\,\mathcal{L}_{P D E}+\lambda_{B C}\,\mathcal{L}_{B C}+\lambda_{I C}\,\mathcal{L}_{I C},
    \label{eq: loss}
\end{gathered}
\end{equation}
where $\lambda_{PDE}$, $\lambda_{BC}$ and $\lambda_{IC}$ are weights assigned to the losses $\mathcal L_{PDE}$, $\mathcal L_{BC}$ and $\mathcal L_{IC}$ respectively.
During training, grid points are sampled within the spatial–temporal domain, ensuring explicit inclusion of boundary points and initial points for enforcing conditions at $\partial \Omega$ and $t=0$\footnote{Grid points at $t=0$ used to impose the initial condition}. The sampled points $\{\mathbf{x}_i, t_i\}_{i=1}^{n_1}, \bm x\in\Omega, t\in (0,T]$ are referred to as collocation points. Explicit output at collocation points is not required, although it can be included if available. For supervised cases where labeled data exists, an additional data-loss term may be incorporated into $\mathcal L_{Total}$. 


Since the total loss function $\mathcal{L}_{\text{Total}}$ involves derivatives of the network $\mathcal{U}(\cdot; \bm{w})$, their computation requires backpropagation through the Variable Spiking Neuron (VSN) layers embedded in the architecture. A conventional approach to address this challenge is the use of surrogate gradients, where the discontinuous dynamics of VSNs are approximated by smooth surrogate functions whose gradients are then employed for optimization. However, in the present context, the network $\mathcal{U}(\cdot; \bm{w})$ directly represents the solution of the governing PDE. Consequently, introducing surrogate gradients would distort the only available reference signal for learning the PDE, potentially leading to inaccuracies and instability during training.
To overcome this limitation, we adopt an upscaling-theory-inspired stochastic projection method for computing the gradients of the NeuroPINN architecture. In this framework, the gradient of the field variable at a given spatial location is estimated using the values observed at its surrounding neighborhood points. The central assumption is that the variation of the field variable $\bm{u}(\bm{x})$ along the neighborhood boundary can be distinctly characterized, while the variations within the neighborhood may be modeled as a stochastic process. This separation allows the boundary point evaluations to be interpreted as macroscopic-scale information, whereas the fluctuations within the neighborhood are treated as microscopic-scale effects.

Considering, $\bm u(\bm z)$ to be a measurement at $\bm z\neq\bm x$, we can write,
\begin{equation}
    \bm u(\bm z)=\bm u(\bm x)+\Delta \bm \eta,
\end{equation}
where $\Delta \bm \eta$ is a zero mean noise, that accounts for fluctuations at the microscopic level. In a differential form, parameterized over time $t$, the above equation takes the following form,
\begin{equation}
    d(\bm u(\bm z_t)-\bm u(\bm x_t))=d\bm \eta_t.
\end{equation}
After sampling information at the micro-scale, at time $t$, the noisy observation at the macroscopic level can thus be written in the form of a stochastic differential equation as,
\begin{equation}
    d \bm Z_t=\bm h\left(\bm x_t, \bm z_t\right) d t+\sigma d \bm W_t,
\end{equation}
where $\bm Z_t$ and $\bm W_t$ represent Brownian motion and $\sigma d\bm W_t$ represents noise based on the microscopic scale. $\bm h\left(\bm x_t, \bm z_t\right)$ returns the difference between the field variable at two macroscopic scale points. 
Now, the field variable $\bm u(\bm x)$ is characterized through a conditional distribution $\pi_t(\bm u)$ in order to introduce microscopically informed spatial variation as,
\begin{equation}
    \pi_t(u) = E_{\mathcal{P}}\left[\bm u(\bm x) \mid \mathcal{F}_{\mathbf{t}}\right],
\end{equation}
where $\mathcal{F}_t$ represents a increasing sigma algebra of $\bm Z_t$ till time $t$. Using the Kallianpur–Striebel formula and a Bayesian filtering approach, the conditional distribution of a continuous and twice differentiable function $\bm \phi$ can be written as,  
\begin{equation}
    \pi_t(\bm \phi) = \frac{E_Q\left[\phi_t \Lambda_t \mid \mathcal{F}_t\right]}{E_Q\left[1 \Lambda_t \mid \mathcal{F}_t\right]},
\end{equation}
where $\Lambda_t$ is the Radon–Nikodym derivative, and $E_Q$ represents expectation under a different probability measure $Q$, which is chosen such that $\bm Z_t$ behaves as a drift-free Brownian motion. Now taking $\bm \phi = \bm z-\bm x$, the evolution equation for $\pi_t(\bm\phi)$ can be written as,  
\begin{equation}
    d \pi_t(\bm \phi) = d \pi_t(\bm z - \bm x) = \left(\pi_t\left((\bm z - \bm x) \bm h^T\right) - \pi_t(\bm h)^T \pi_t(\bm z - \bm x)\right) \cdot (\bm \sigma \bm \sigma^T)^{-1} \left(d\bm Z_t - \pi_t(\bm h) dt\right).
\end{equation}
The time evolution for the above equation starting from an initial time $t_0$ can be written as,
\begin{equation}
    (z_t - x_t) = (z_{t_0} - x_{t_0}) + \int_{t_0}^{t} \left( \pi_s\left( (\bm z - \bm x) \bm h^T \right) - \pi_s(\bm h)^T \pi_s(\bm z - \bm x) \right) \cdot (\bm \sigma \bm \sigma^T)^{-1} d\bm Z_s.
\end{equation}
It is important to note that the correction term $\bm{h}$ vanishes (i.e., $\bm{h} = \bm{0}$) whenever the difference between the state variables $\bm{z}_t$ and $\bm{x}_t$ remains within the prescribed characteristic distance.
By assuming local stationarity and taking an ensemble average over neighborhood points, we obtain a nonlocal estimate of the gradient,  
\begin{equation}
    G(X = \bar{X}) = \frac{\partial u}{\partial X} = \frac{\frac{1}{N_t} \sum_{i=1}^{N_t}(u-\bar{u})\left(X_i-\bar{X}\right)^T}{\frac{1}{N_t} \sum_{i=1}^{N_t}\left(X_i-\bar{X}\right)\left(X_i-\bar{X}\right)^T}.
    \label{eq: gradients}
\end{equation}
The required gradients, denoted by $G(\cdot)$, at a given point $\bm{x}_p = {\bm{x}, t}$ can be evaluated by utilizing the information from its neighborhood. Specifically, the neighborhood consists of points ${\bm{x}^{(n)}i}{i=1}^{N_t}$, which are distributed within a prescribed critical distance from $\bm{x}_p$. The gradient computation at $\bm{x}_p$ is then expressed as
\begin{equation}
    \mathbf{G}(\bm x=\bm x_p)=\frac{\partial \mathcal U(\bm x_p, \bm w)}{\partial \bm x}=\dfrac{\sum\limits_{i=1}^{N_t}\left(\mathcal U\left(\bm x^{(n)}_i, \bm w\right)-\mathcal U(\bm x_p, \bm w)\right)\left(\bm x^{(n)}_i-\bm{x_p}\right)^T}{ \sum\limits_{i=1}^{N_t}\left(\bm x^{(n)}_i-\bm x_p \right)\left(\bm x^{(n)}_i-\bm x_p\right)^T}.
    \label{eq: gradients sp}
\end{equation}
For more details on the stochastic projection method, readers are referred to \cite{nowruzpour2019derivative,navaneeth2023stochastic}.
\begin{figure}[ht!]
    \centering
    \includegraphics[width=\linewidth]{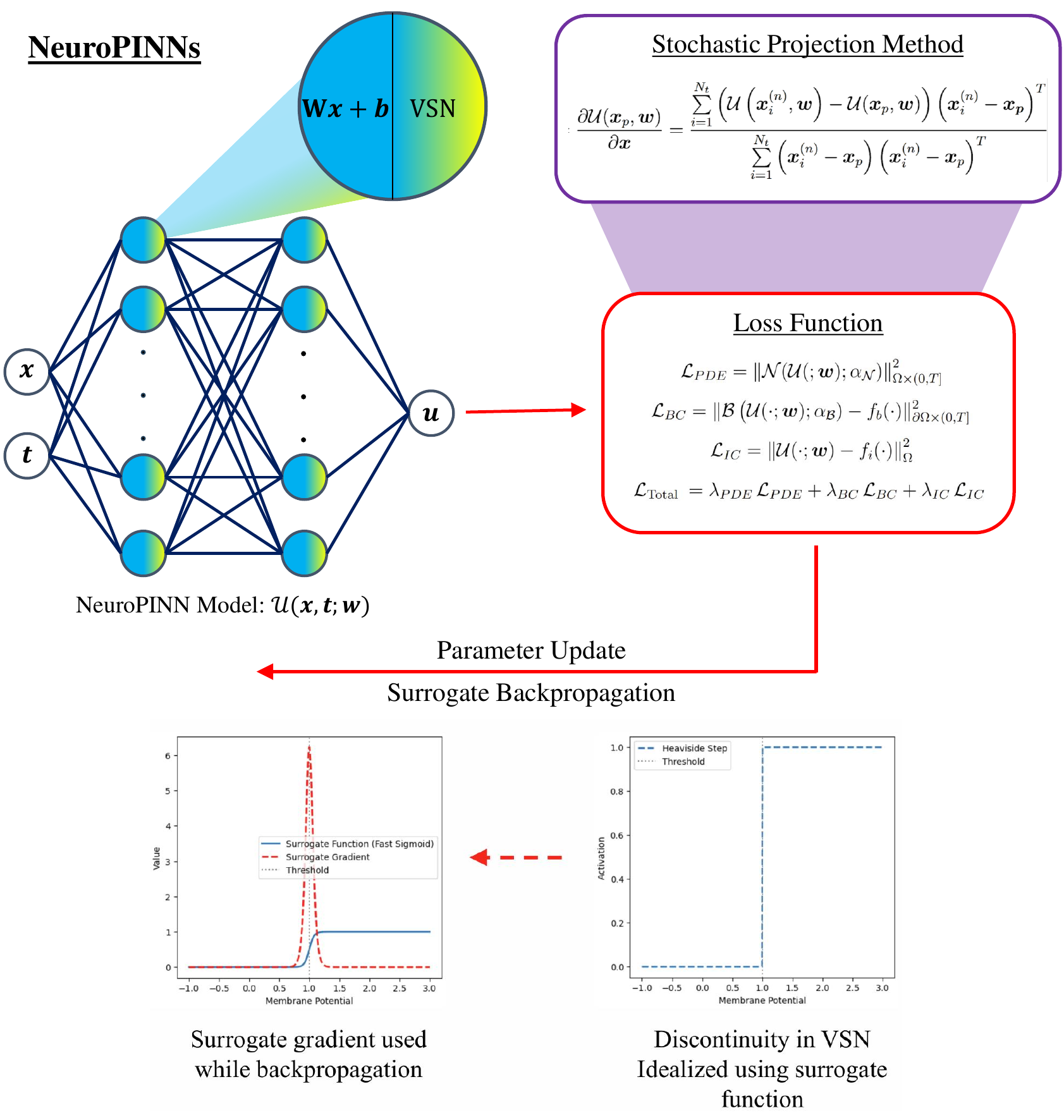}
    \caption{Schematic for the proposed NeuroPINNs.}
    \label{fig: spNeuroPINN flow}
\end{figure}
\begin{algorithm}[ht!]
\caption{Training algorithm for NeuroPINNs}
\label{algo}
\textbf{Requirements:} The residual form of the PDE to be learned, along with its boundary conditions and initial conditions. Sample collocation points, $\{\bm x_i, t_i\}_{i = 1}^{n_1}$, boundary points $\{\bm x_i, t_i\}_{i = 1}^{n_2}$ and the initial points $\{\bm x_i, 0\}_{i = 1}^{n_3}$. A base network architecture with initialized trainable parameters $\bm w$ that will reflect the solution of the PDE. Neighborhood points for the various sampled points for computing the required gradients using stochastic projection.
\begin{algorithmic}[1]
\For{$e = 1$ to number of epochs}
    \State Get network prediction for the collocation points, boundary points, and the initial points.
    \State Get the first order derivatives of outputs with respect to the inputs using Eq. \eqref{eq: gradients sp}.
    \State Higher order gradients can also be computed using the Eq. \eqref{eq: gradients sp}, as applied to gradients of the previous order.
    \State Compute the loss function $\mathcal{L}_{\text{Total}}$ as shown in Eq. \eqref{eq: loss}.
    \State Compute gradients of loss function with respect to the trainable parameters $\bm \theta$ using surrogate backpropagation.
    \State Update trainable parameters using a suitable optimization algorithm like Adam optimizer.
\EndFor
\end{algorithmic}
\textbf{Output}: Trained NeuroPINN network with optimized trainable parameters.
\end{algorithm}
An algorithm for training the proposed NeuroPINNs is given in Algorithm \ref{algo}. A schematic for the proposed NeuroPINNs is given in Fig. \ref{fig: spNeuroPINN flow}.
Since the VSNs in NeuroPINNs work with spike trains, potentially containing multiple STSs, the inputs being passed may be encoded using techniques like rate encoding or time encoding. However, we propose using direct inputs at all STSs in order to preserve the input information and maintain the overall accuracy of the results. In the event of multiple STSs being considered in the NeuroPINN architecture, the information after the last VSN layer will need to be collected, and its mean will be used for processing in further layers of the network. 
\subsection{Energy Consumption}
The primary motivation for employing spiking neurons lies in their potential for energy efficiency, particularly when implemented on event-driven hardware. To illustrate how sparse communication can yield energy savings, consider two densely connected layers. Let the first layer contain $n_i$ nodes and the second layer contain $n_j$ nodes. The $n_i$ outputs from the first layer are first transmitted through a spiking layer before being propagated to the second layer. In a conventional densely connected setting, synaptic operations consist of multiplying the weight matrix with the incoming inputs, resulting in a computational cost of $n_i n_j$ operations. The energy expenditure associated with computing the inputs themselves is disregarded here, since it scales as $\mathcal{O}(n_i)$ and is strongly dependent on the specific implementation of the activation function.

The energy consumption in synaptic operations is predominantly determined by fixed mathematical procedures, which allows for reliable generalization. As discussed in \cite{davidson2021comparison}, four fundamental operations contribute to this cost: (i) reading the neuron state, (ii) performing a multiplication, (iii) performing an addition, and (iv) writing the neuron state. Based on post-layout energy estimation data from SpiNNaker2, \cite{davidson2021comparison} reports that both multiplication and reading the neuron state require $5E$ units of energy, where $E$ denotes the energy required for a single addition operation (also equal to the energy cost of writing the neuron state). Consequently, for an artificial neural network (ANN) employing continuous activations with 100\% spiking activity, the total energy consumption in synaptic operations can be expressed as
\begin{equation}
    n_1n_2(5E+5E+E+E) = 12En_1n_2.
\end{equation}
 
Extending this analysis to variable spiking neurons (VSNs), the presence of variable-amplitude spikes implies that all of the aforementioned energy-consuming operations are required. However, these costs are incurred only when a spike occurs. Accordingly, the energy $E_{\text{VSN-Syn}}$ consumed in synaptic operations, when the input originates from a VSN layer, can be expressed as,
\begin{equation}
    E_{VSN-Syn} = (n_1 N_{\text{avg spikes}})n_2(5E+5E+E+E) = 12En_1n_2N_{\text{avg spikes}},
    \label{eqn: energy}
\end{equation}
where $N_{\text{avg spikes}}$ denotes the average number of spikes generated in the VSN layer. Comparing this with the synaptic energy consumption of an ANN, it follows that energy savings will occur provided the average spike count remains below 100\%. Since the energy expenditure in an event-driven computing paradigm scales directly with spike activity, the subsequent discussion will use spiking activity as a proxy for energy savings. This observation aligns with prior reports on the energy efficiency of spiking neurons \cite{davidson2021comparison,dampfhoffer2022snns,lemaire2022analytical}. It is worth noting, however, that while vanilla spiking neurons consume less energy—owing to the elimination of the multiplication operation—their utility is limited by reduced accuracy, particularly in regression-oriented tasks.
\section{Numerical Illustrations}\label{section: numerical}
In this section, we present four examples, along with a three-dimensional application in linear elastic micromechanics, to demonstrate the performance of the proposed NeuroPINN framework. The first two examples (E-I and E-II) address time-dependent, one-dimensional problems: the Burgers equation and the heat conduction equation, respectively. The next two examples (E-III and E-IV) focus on the two-dimensional Poisson equation, solved on L-shaped and star-shaped domains. To benchmark performance, we compare NeuroPINNs with vanilla SP-PINNs and CPINNs, both of which employ spike-based processing (SP) to compute the loss function. When surrogate backpropagation is adopted for gradient computation in the loss evaluation, we denote the corresponding architecture with the prefix SB, for example, SB-NeuroPINN.
In addition, we consider a three-dimensional application to linear elastic micromechanics, which involves learning stress responses in synthetic polycrystalline microstructures. The full details of this case study are presented in a dedicated subsection.

\begin{table}[ht!]
\centering
\caption{Network details for various examples.  $n_l$ here represents the number of layers used in the deep learning model of various examples. The configuration here shows the arrangement of feed-forward layers and the number of nodes in each layer. $\sigma(\cdot)$ is the activation used between various layers when using vanilla artificial neurons.}
\label{table:arch_details}
\begin{tabular}{lccccc}
\toprule
 Example & $n_l$ & Configuration & $\sigma(\cdot)$ & $n_c$ & $n_b+n_i$ \\
 \midrule
 E-I, Burgers Equation & 5 & 2-40-80-40-1 & tanh & 2601 & 265+135 \\
 E-II, Heat Equation & 5 &  2-40-80-40-1 &  tanh & 2601 & 265+135 \\
 E-III, Poissons Equation - L Shaped & 5 &  2-40-80-40-1 & ELU & 1935 & 600+0 \\
 E-IV, Poissons Equation - Star Shaped & 5 &  2-40-120-40-1 &  tanh & 4595 & 600+0 \\
 \bottomrule
\end{tabular}
\end{table}
The neural network configurations employed in the different examples are summarized in Table \ref{table:arch_details}. The sequence of numbers specifies the number of nodes in the input layer (first entry), hidden layers, and the output layer (last entry). Each hidden layer is equipped with an activation function, $\sigma(\cdot)$. In the case of networks incorporating VSNs, namely NeuroPINNs and SB-NeuroPINNs, the first and second activation layers are replaced by VSN layers. The number of VSNs in each layer matches the number of nodes in the corresponding wrapped activation layer. Table \ref{table:arch_details} also provides the number of collocation points ($n_c$), boundary points ($n_b$), and initial points ($n_i$) used in each example. Training of the NeuroPINN architectures is performed using the Adam optimizer, with a single STS employed unless stated otherwise. By contrast, CPINNs utilize 256 STSs for training. Finally, note that $\sigma(\cdot)$ in Table \ref{table:arch_details} also serves as the continuous activation within the VSN. Note that the network details for the three-dimensional micromechanics application are provided separately in its subsection, given its different style of architecture and dataset requirements
\subsection{E-I, Burgers Equation}
In the first example, we consider the one-dimensional, time-dependent Burgers’ equation. This equation is widely regarded as a canonical model, with applications in shock wave analysis, traffic flow modeling, and gas dynamics. The governing equation is given as,
\begin{equation}
\begin{gathered}
    \dfrac{\partial u(x,t)}{\partial t}+u(x,t)\dfrac{\partial u(x,t)}{\partial x} = \dfrac{\partial^2 u(x,t)}{\partial x^2}, \,\,x\in[0,1],\,\,t\in(0,1],\\
    u(x,0) = \sin\left(\dfrac{\pi x}{l}\right),\\
    u(0,t) = u(1,t) = 0.
\end{gathered}
\end{equation}
\begin{table}[ht!]
    \centering
    \caption{Percentage relative $L^2$ errors observed when comparing ground truth and network predictions. DNC here refers to Did Not Converge.}
    \label{table:l2_errors}
    \begin{tabular}{l>{\centering\arraybackslash}m{2cm}>{\centering\arraybackslash}m{2cm}>{\centering\arraybackslash}m{2cm}>{\centering\arraybackslash}m{2cm}}
        \toprule
        \multirow{2}{*}{\textbf{Model}} & \multicolumn{4}{c}{\textbf{Equation}} \\ 
        \cmidrule{2-5}
        & \textbf{E-I} & \textbf{E-II} & \textbf{E-III} & \textbf{E-IV} \\ 
        \midrule
        SP-PINN  & 0.25 & 0.46 & 3.67 & 0.26\\ 
        NeuroPINN  & 1.04 & 1.27 & 2.70 & 0.34\\ 
        SB-NeuroPINN  & 6.70 & 1.96 & DNC  & 0.87\\
        CPINN  & 11.87 & 13.75 & DNC & 13.06\\ 
        \hline
    \end{tabular}
\end{table}

Table \ref{table:l2_errors} presents the percentage relative $L^2$ errors obtained using different deep learning models. For this example, we perform predictions on a grid of resolution $201 \times 201$. We observe that the NeuroPINN achieves lower errors compared to the SB-NeuroPINN, with error levels much closer to those obtained using the vanilla SP-PINN. This outcome is consistent with our expectation that employing surrogate gradients in the loss computation introduces inaccuracies. Furthermore, the NeuroPINN also yields better accuracy than the CPINN. This result is again anticipated, since in CPINNs the trained vanilla PINN architectures are mapped to the spiking domain through additional assumptions and approximations, which inevitably introduce errors.
\begin{figure}[ht!]
    \centering
    \begin{subfigure}{0.245\linewidth}
        \centering
        \includegraphics[width=\linewidth]{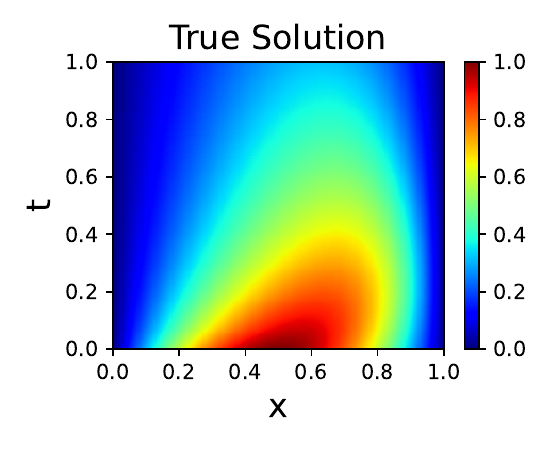}
        \caption{Ground Truth}
    \end{subfigure}


    \begin{subfigure}{0.49\linewidth}
        \centering
        \includegraphics[width=\linewidth]{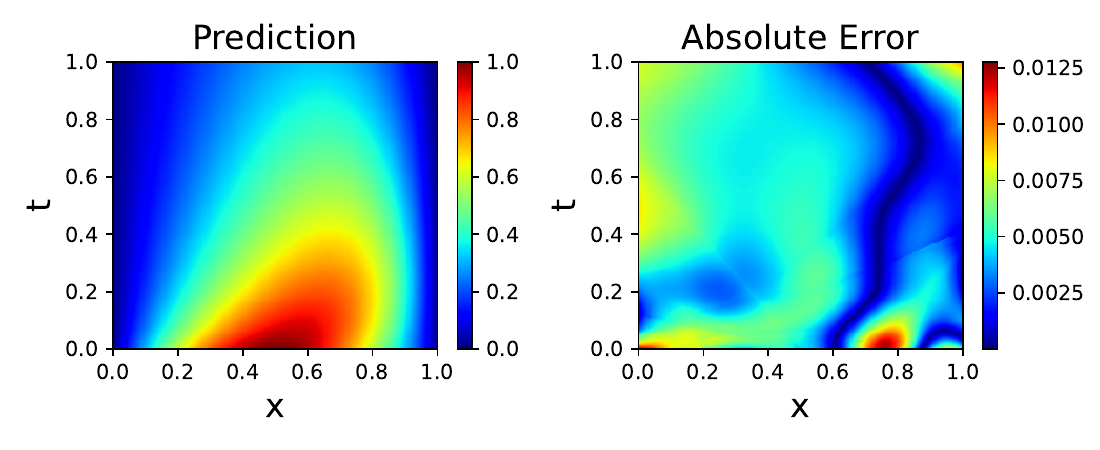}
        \caption{NeuroPINN Prediction}
    \end{subfigure}
    \begin{subfigure}{0.49\linewidth}
        \centering
        \includegraphics[width=\linewidth]{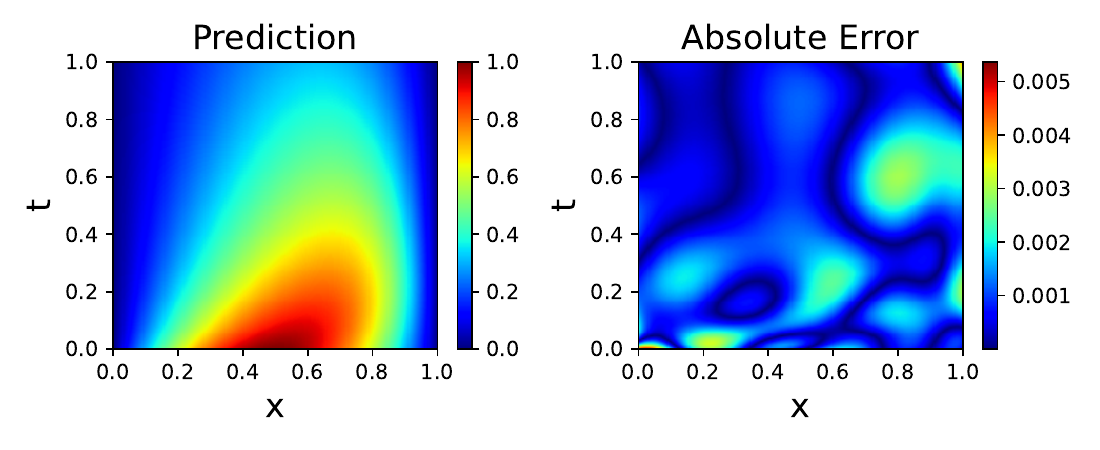}
        \caption{SP-PINN Prediction}
    \end{subfigure}


    \begin{subfigure}{0.49\linewidth}
        \centering
        \includegraphics[width=\linewidth]{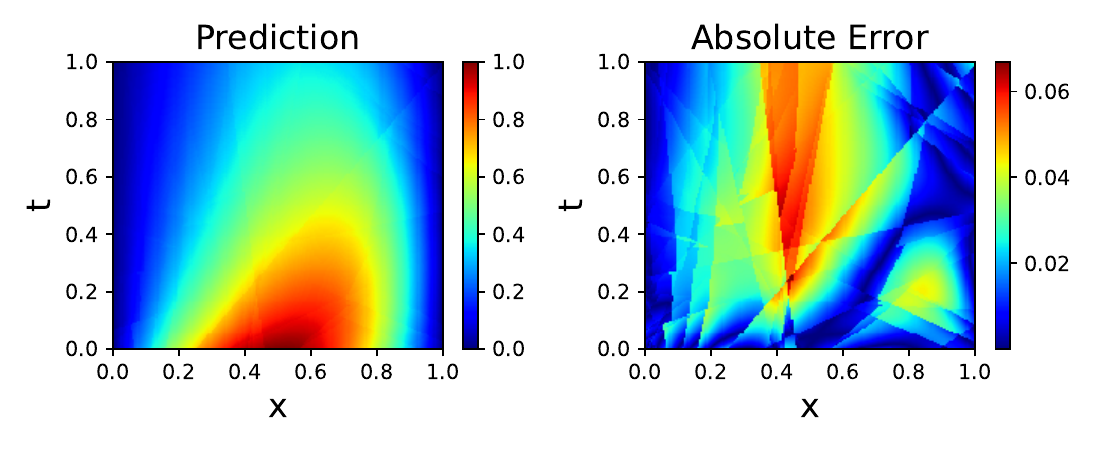}
        \caption{SB-NeuroPINN Prediction}
    \end{subfigure}
    \begin{subfigure}{0.49\linewidth}
        \centering
        \includegraphics[width=\linewidth]{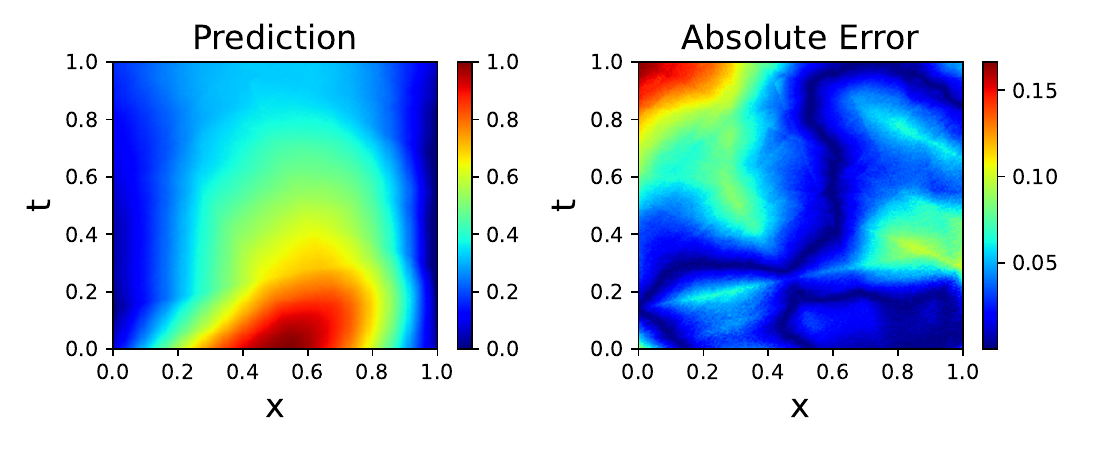}
        \caption{CPINN Prediction}
    \end{subfigure}

    \caption{Ground truth compared against various network predictions in E-I.}
    \label{fig:burger_pred_true_plots}
\end{figure}
Figure \ref{fig:burger_pred_true_plots} provides a visual validation of the results reported in Table \ref{table:l2_errors}. We observe that the predictions obtained using NeuroPINN align closely with the ground truth, further confirming the accuracy of the approach.
\subsection{E-II, Heat Conduction Equation}
In the second example, we consider the one-dimensional time-dependent heat conduction equation, which is solved using a range of physics-informed deep learning models. This equation is of broad relevance, with applications spanning environmental science (climate studies), medical science (heat transfer in biological tissue), and astrophysics (planetary thermal processes), among others. The governing equation is formulated as follows:
\begin{equation}
\begin{gathered}
    \dfrac{\partial u(x,t)}{\partial t}=\dfrac{\partial^2 u(x,t)}{\partial x^2}+100\sin\left(\dfrac{\pi x}{l}\right), \,\,x\in[0,1],\,\,t\in(0,1],\\
    u(x,0) = 0,\\
    u(0,t) = u(1,t) = 0.
\end{gathered}
\end{equation}
\begin{figure}[ht!]
    \centering
    \begin{subfigure}{0.245\linewidth}
        \centering
        \includegraphics[width=\linewidth]{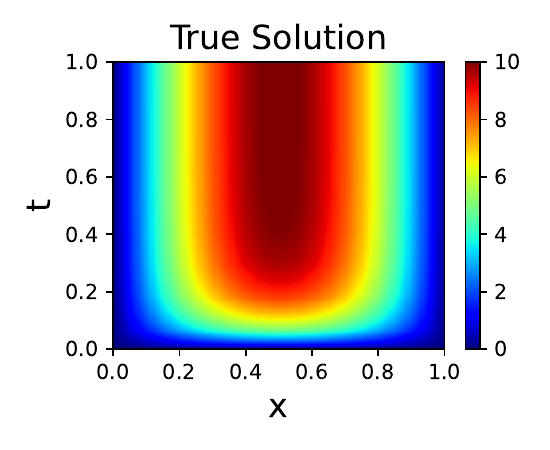}
        \caption{Ground Truth}
    \end{subfigure}


    \begin{subfigure}{0.49\linewidth}
        \centering
        \includegraphics[width=\linewidth]{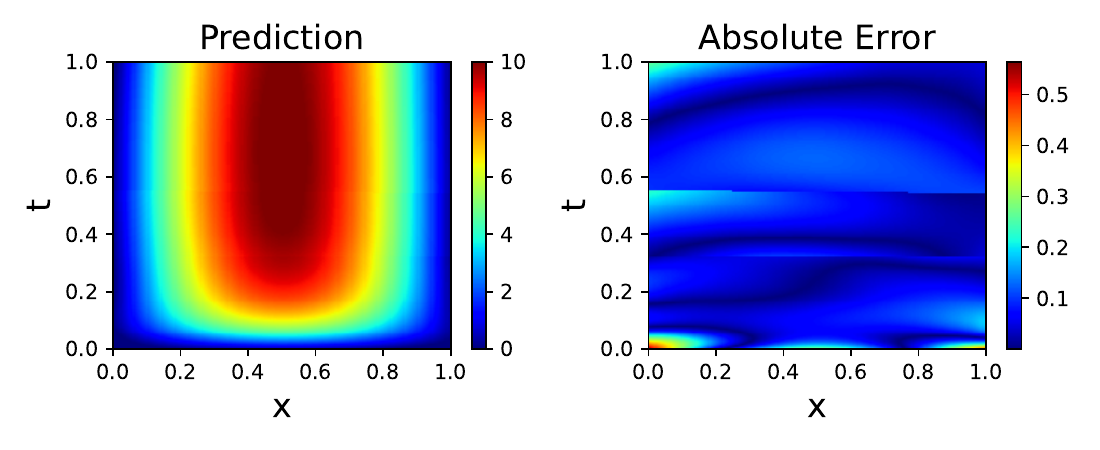}
        \caption{NeuroPINN Prediction}
    \end{subfigure}
    \begin{subfigure}{0.49\linewidth}
        \centering
        \includegraphics[width=\linewidth]{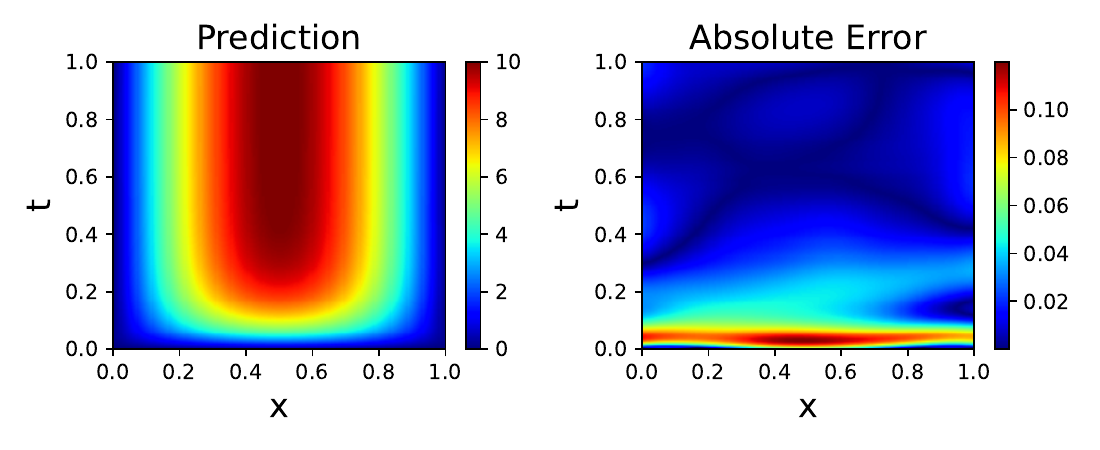}
        \caption{SP-PINN Prediction}
    \end{subfigure}


    \begin{subfigure}{0.49\linewidth}
        \centering
        \includegraphics[width=\linewidth]{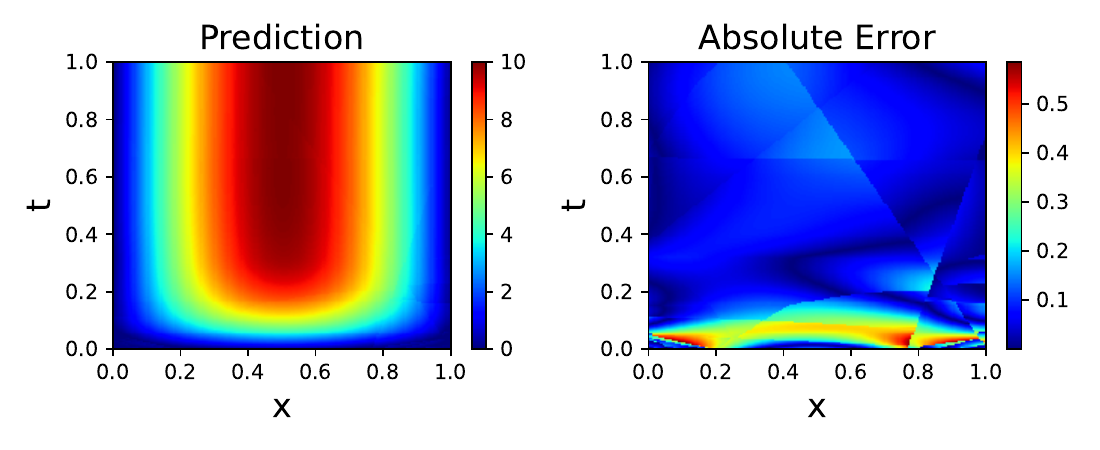}
        \caption{SB-NeuroPINN Prediction}
    \end{subfigure}
    \begin{subfigure}{0.49\linewidth}
        \centering
        \includegraphics[width=\linewidth]{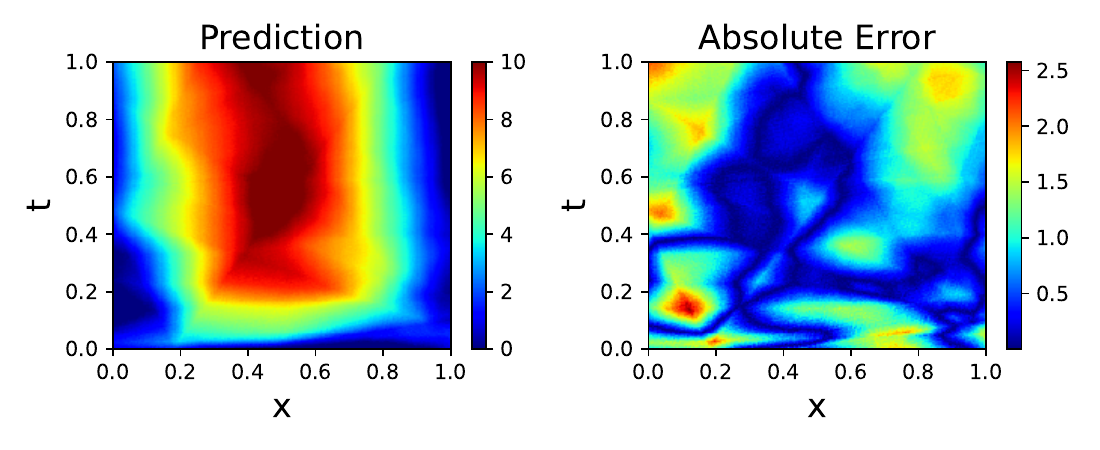}
        \caption{CPINN Prediction}
    \end{subfigure}

    \caption{Ground truth compared against various network predictions in E-II.}
    \label{fig:heat_pred_true_plots}
\end{figure}
Table \ref{table:l2_errors} and Fig. \ref{fig:heat_pred_true_plots} exhibit a trend consistent with the observations from the previous example. The NeuroPINN demonstrates superior performance compared to both SB-NeuroPINN and CPINN models, with its predictions showing close agreement with the ground truth. This outcome is expected: the reliance of SB-NeuroPINN on surrogate gradients during loss function computation introduces systematic errors, while the CPINN architecture inherits inaccuracies from the approximations made when converting trained vanilla PINNs into the spiking domain. By contrast, the NeuroPINN framework eliminates these limitations, thereby achieving lower error magnitudes. The predictions were evaluated on a grid with a resolution of $201\times201$.
\subsection{E-III, Poisson Equation on L-shaped Domain}
In the third example, we solve the two-dimensional Poisson equation. Unlike the previous cases, the solution domain considered here is irregular, taking the form of an L-shape. The Poisson equation plays a central role in a wide range of applications, including electrostatics, astrophysics, and fluid mechanics. The specific form of the equation under consideration is given as,
\begin{equation}
\begin{gathered}
    \dfrac{\partial^2 u(x,y)}{\partial x^2}+\dfrac{\partial^2 u(x,y)}{\partial y^2}=1, \,\,x\in[-1,1],\,\,y\in[-1,1],\\
\end{gathered}
\end{equation}
\begin{figure}[ht!]
    \centering
    \begin{subfigure}{0.245\linewidth}
        \centering
        \includegraphics[width=\linewidth]{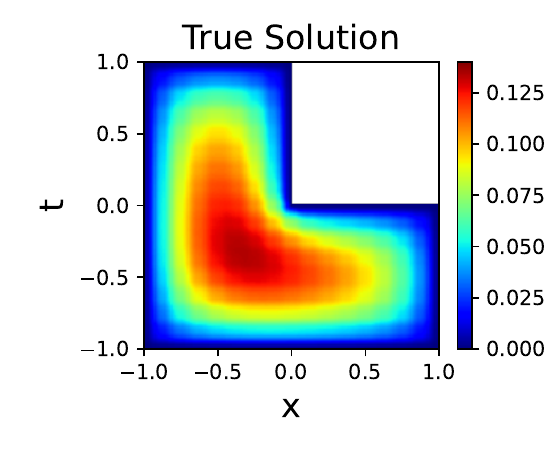}
        \caption{Ground Truth}
    \end{subfigure}


    \begin{subfigure}{0.49\linewidth}
        \centering
        \includegraphics[width=\linewidth]{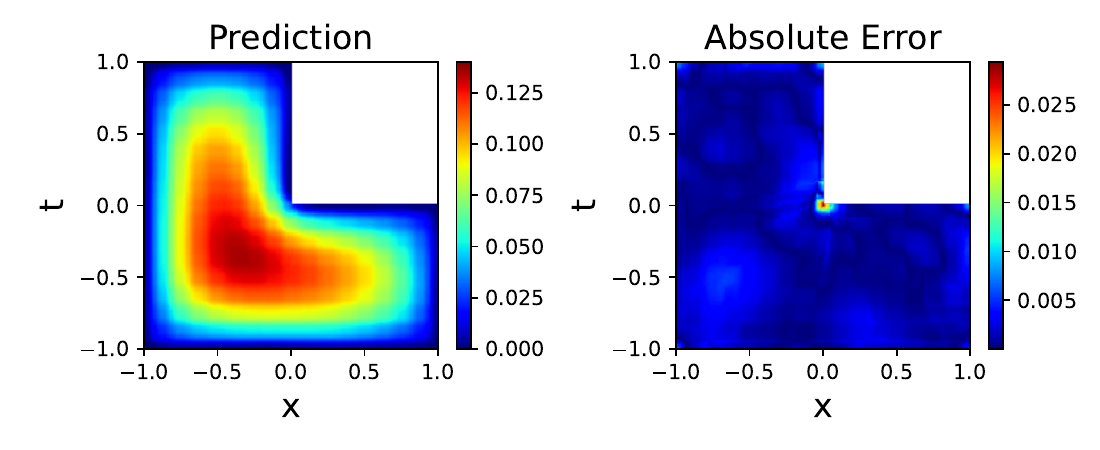}
        \caption{NeuroPINN Prediction}
    \end{subfigure}
    \begin{subfigure}{0.49\linewidth}
        \centering
        \includegraphics[width=\linewidth]{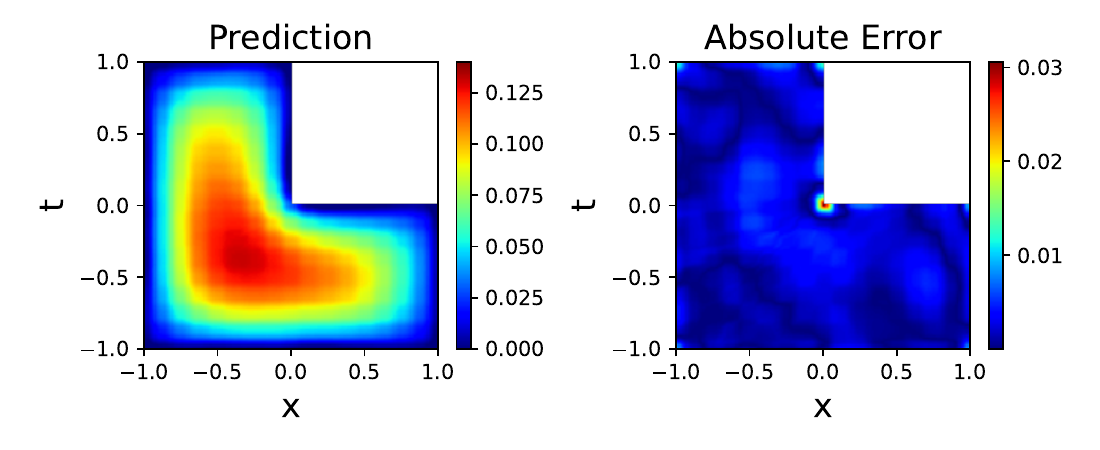}
        \caption{SP-PINN Prediction}
    \end{subfigure}


    \caption{Ground truth compared against various network predictions in E-III.}
    \label{fig:poisson_pred_true_plots}
\end{figure}
Table \ref{table:l2_errors} presents the relative $L^2$ errors for the irregular-domain Poisson problem. The results clearly indicate that NeuroPINN demonstrates superior performance compared to the AD-PINN model, achieving significantly lower errors. In contrast, both SB-NeuroPINN and CPINN models failed to converge, highlighting the challenges these architectures face when applied to irregular domains. This finding underscores the robustness of the NeuroPINN framework in handling complex geometries where other spiking or conversion-based approaches struggle. Furthermore, Fig. \ref{fig:poisson_pred_true_plots} provides a visual confirmation of this trend, showing that NeuroPINN predictions exhibit a close match with the ground truth solution, thereby reinforcing the quantitative results reported in Table \ref{table:l2_errors}.
\subsection{E-IV, Poisson Equation on Star-shaped Domain}
In the fourth example, we consider the Poisson equation on a star-shaped domain, which introduces an additional level of geometric complexity compared to the earlier L-shaped case. Solving PDEs on such irregular and non-convex domains is of significant interest, as they frequently arise in practical applications such as electrostatics, structural mechanics, and fluid flow in complex geometries. The analytical solution for the equation in the current example is defined as,
\begin{equation}
\begin{gathered}
    u(x,y) = exp(-(2x^2+4y^2))+\dfrac{1}{2}, \,\,x\in[-1.5,1.5],\,\,y\in[-1.5,1.5].
\end{gathered}
\end{equation}
\begin{figure}[ht!]
    \centering
    \begin{subfigure}{0.245\linewidth}
        \centering
        \includegraphics[width=\linewidth]{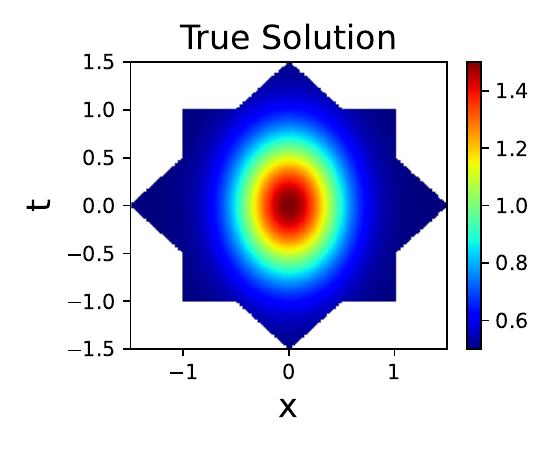}
        \caption{Ground Truth}
    \end{subfigure}


    \begin{subfigure}{0.49\linewidth}
        \centering
        \includegraphics[width=\linewidth]{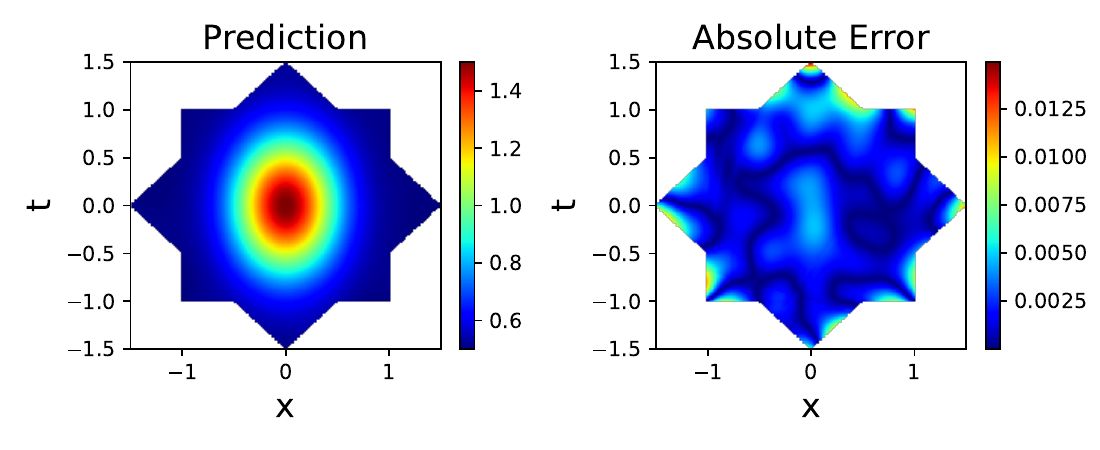}
        \caption{NeuroPINN Prediction}
    \end{subfigure}
    \begin{subfigure}{0.49\linewidth}
        \centering
        \includegraphics[width=\linewidth]{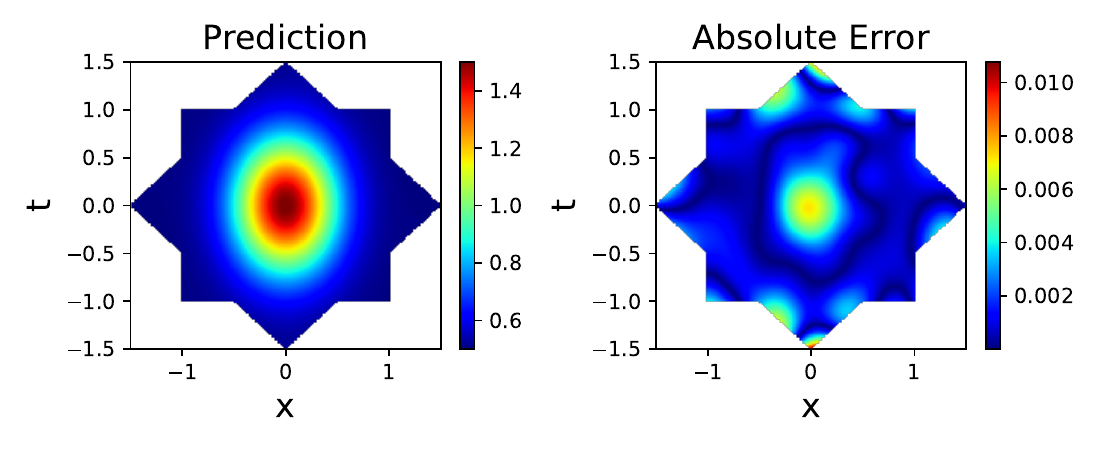}
        \caption{SP-PINN Prediction}
    \end{subfigure}


    \begin{subfigure}{0.49\linewidth}
        \centering
        \includegraphics[width=\linewidth]{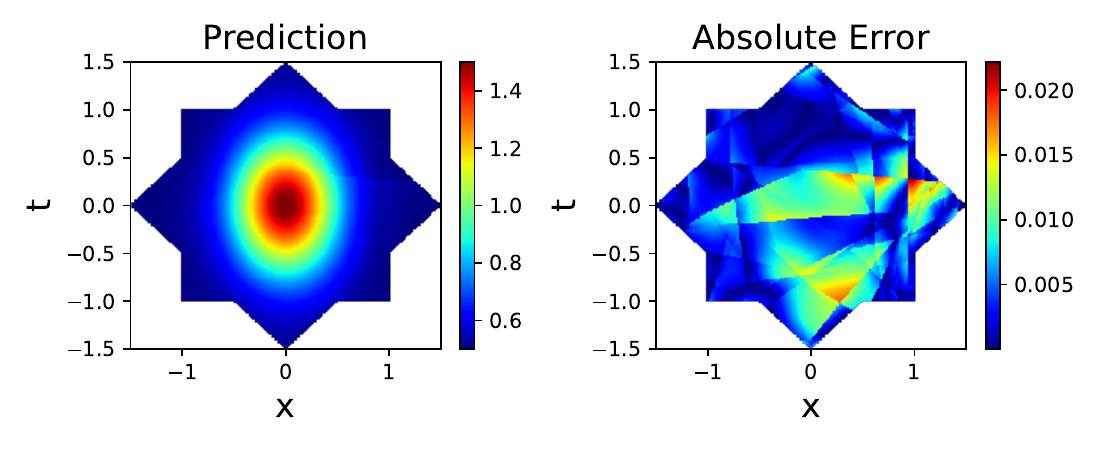}
        \caption{SB-NeuroPINN Prediction}
    \end{subfigure}
    \begin{subfigure}{0.49\linewidth}
        \centering
        \includegraphics[width=\linewidth]{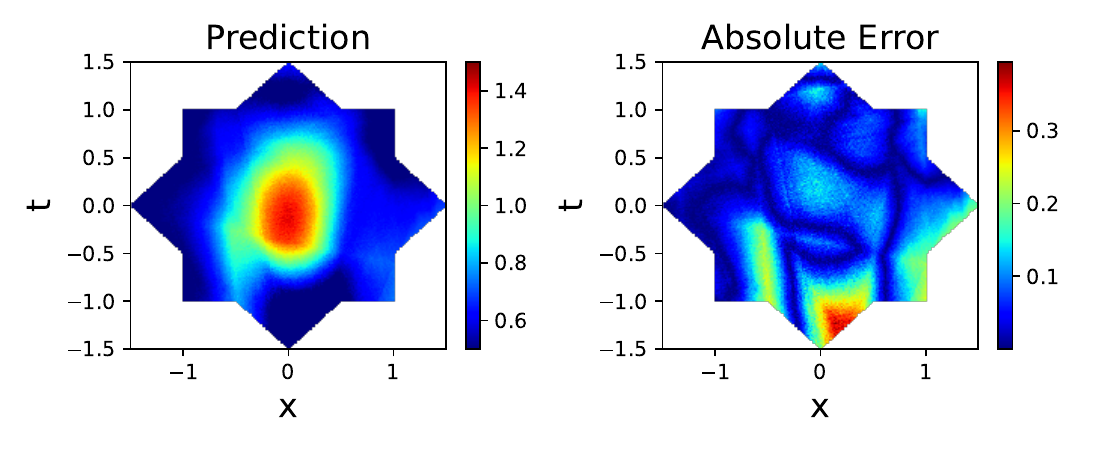}
        \caption{CPINN Prediction}
    \end{subfigure}

    \caption{Ground truth compared against various network predictions in E-IV.}
    \label{fig:poisson_star_pred_true_plots}
\end{figure}
Table \ref{table:l2_errors} and Fig. \ref{fig:poisson_star_pred_true_plots} demonstrate a consistent trend with those observed in Examples I and II. The NeuroPINN model achieves superior accuracy compared to both SB-NeuroPINN and CPINN, with its predictions closely matching the ground truth solution. It is worth noting that while additional optimization of the calibration process could potentially improve the performance of CPINNs, the results clearly highlight the markedly better capability of NeuroPINN networks in learning and generalizing across different classes of PDEs. Furthermore, we conducted additional experiments using LIF-neuron-augmented PINN networks with a single spiking time step (STS). However, these attempts failed to converge to the ground truth solution across the test cases considered, thereby emphasizing the robustness and reliability of the proposed NeuroPINN approach.
\subsection{Spiking Activity and Multiple STS Performance: NeuroPINN Models}
Since the primary motivation for employing NeuroPINN networks lies in their potential for energy savings—enabled by their ability to perform sparse communication, it is essential to examine the spiking activity of the VSN layers across different NeuroPINN models.

\begin{figure}[ht!]
\centering
\includegraphics[width=0.6\linewidth]{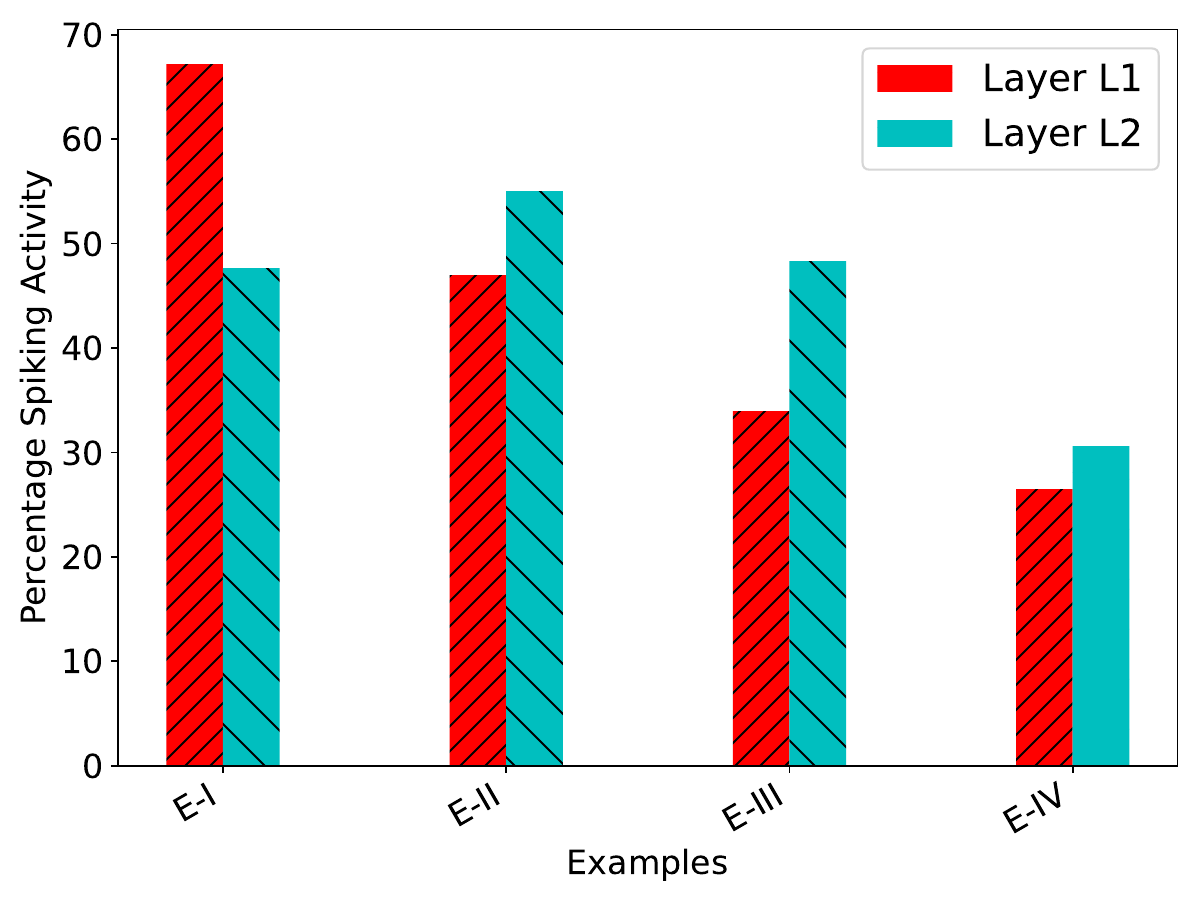}
\caption{Spiking activity observed in NeuroPINN networks across various examples.}
\label{fig:spk_act}
\end{figure}
Figure \ref{fig:spk_act} illustrates the spiking activity recorded in the two VSN layers of NeuroPINN models. In all examples, the observed activity remains well below 100\%, which indicates significant potential for energy savings when deployed on neuromorphic hardware. Following the framework proposed in \cite{davidson2021comparison}, this spiking activity can be used to provide an initial estimate of the energy savings in synaptic operations within densely connected layers. To quantify this, we define an energy ratio $E_r$, which is the ratio of energy consumed in synaptic operations based on the observed spiking activity to that consumed under 100\% spiking activity in all densely connected layers. The calculated values of $E_r$ are 0.58, 0.52, 0.42, and 0.29 for the four examples, respectively. These results suggest that synaptic operations require approximately half the original energy in E-I and only about one-third in E-IV. This gap is likely to be even larger in practice, given that neuromorphic chips are intrinsically more power-efficient than GPUs and CPUs typically used for running deep learning models \cite{parpart2023implementing}.

\begin{table}[ht!] \centering \caption{Percentage relative $L^2$ errors when using two spike time steps. Cumulative spiking activity across both spike time steps in different spiking layers is also shown. SA here represents Spiking Activity.} \label{table:2sts} \begin{tabular}{l>{\centering\arraybackslash}m{2cm}>{\centering\arraybackslash}m{2cm}>{\centering\arraybackslash}m{2cm}>{\centering\arraybackslash}m{2cm}} \toprule \multirow{2}{*}{\textbf{ }} & \multicolumn{4}{c}{\textbf{Equation}} \\ \cmidrule{2-5} & \textbf{E-I} & \textbf{E-II} & \textbf{E-III} & \textbf{E-IV} \\ \midrule Error & 0.49 & 1.06 & 2.67 & 0.22\\ SA-L1 & 0.50 & 0.70 & 0.38 & 0.35\\ SA-L2 & 0.72 & 0.55 & 0.27 & 0.14\\ \hline 
\end{tabular} 
\end{table}
To further evaluate the benefits of multiple STS, we conducted a case study where NeuroPINN models for all examples were run with two STS. The results, presented in Table \ref{table:2sts}, clearly indicate that the use of multiple STS reduces the error, while the cumulative spiking activity across both STS remains well below 100
\subsection{Application of NeuroPINN for Linear Elastic Micromechanics in Three Dimensions} 
In this example, we utilize the NeuroPINN model for solving linear elastic micromechanics in three dimensions. Synthetic polycrystalline microstructures are generated on a cubic domain $\Omega \subset \mathbb{R}^3$, representing a periodic cell of the material. The grain geometry is constructed by a Voronoi tessellation, which partitions the cubic domain into convex grains, each corresponding to the region of points closest to a given randomly distributed seed point. These seeds are sampled from a Matérn hardcore point process with periodic distance, which enforces a minimum separation radius $r$ between seeds. This prevents unrealistically small grains and leads to more representative morphologies. The tessellation is periodic by construction, and the resulting grains are convex polytopes. Each grain is assigned a crystallographic orientation, sampled independently 
from the uniform distribution on Special Orthogonal Group of degree 3, $\mathrm{SO}(3)$. Orientations are 
represented by unit quaternions, corresponding to points on the 
3-sphere $S^3 \subset \mathbb{R}^4$, in order to avoid singularities in 
rotation parameterizations. 
For the dataset, the cubic domain is discretized into 
$32^3$ voxels. With parameters $\lambda = 15/32^3$ for the point process 
intensity and $r = 5$ for the hardcore radius, the expected number of grains 
per cell is about $15$, with each grain occupying on average roughly 
2200 voxels. This resolution balances the need to capture intra-grain fields 
with computational tractability.

The network architecture consists of a 3D residual convolutional neural network (ResNet) that predicts the periodic part of the displacement field, $\mathbf{u}^*(\mathbf{x}) \in \mathbb{R}^3$, from the local crystallographic orientations. The input tensor has $4$ channels, corresponding to the components of the unit quaternion at each voxel, while the output tensor has $3$ channels, corresponding to the components of $\mathbf{u}^*$. 
This is the trainable part of the network  and is a fully convolutional encoder-decoder architecture with residual connections. Instance normalization and Leaky ReLU activations are used throughout. 
The architecture also has non-trainable layers that enforce the mechanical structure of the problem as follows,
\begin{enumerate}
    \item \textbf{Orientation to stiffness mapping:} 
    Each quaternion $\mathbf{q}(\mathbf{x}) \in S^3$ is converted into a passive rotation matrix 
    $G(\mathbf{q}) \in \mathrm{SO}(3)$. 
    The local stiffness tensor is then obtained by the standard fourth-order tensor rotation,
    \begin{equation}
        \mathbb{C}_{ijkl}(\mathbf{x}) 
        = G_{ip} G_{jq} G_{kr} G_{ls} \, (\mathbb{C}_0)_{pqrs},
    \end{equation}
    where $\mathbb{C}_0$ is the reference crystal stiffness (tetragonal TiAl in this case). 
    \item \textbf{Strain computation:} The predicted displacement field is combined with the applied macroscopic strain tensor $\mathbf{E}$ to form the local strain,
    \[
    \hat{\boldsymbol{\varepsilon}}(x) = \mathbf{E} + \tfrac{1}{2} \left( \nabla \hat{\mathbf{u}}^*(x) + \nabla \hat{\mathbf{u}}^*(x)^\top \right),
    \]
    where spatial gradients are computed by fixed forward finite-difference kernels, implemented as non-trainable $3\times 3\times 3$ convolutions.
    \item \textbf{Stress computation:} The Cauchy stress is then given by
    \[
    \hat{\boldsymbol{\sigma}}(x) = \mathbb{C}(x) : \hat{\boldsymbol{\varepsilon}}(x).
    \]
    \item \textbf{Equilibrium residual:} The divergence of the stress field is obtained using fixed backward finite-difference kernels,
    \[
    \mathbf{r}(x) = \nabla \cdot \hat{\boldsymbol{\sigma}}(x),
    \]
    which serves as the residual in the loss term.
\end{enumerate}
When using NeuroPINN, an architecture similar to above is adopted; however, the activations within the trainable part of the network are replaced by VSN layers, and stochastic projection is used to compute the final loss term. 
The training dataset consists of $5000$ independent realizations of synthetic microstructures. Training is performed with the Adam optimizer, batch size $1$, and initial learning rate $10^{-3}$. The output obtained from the network is further processed to obtain the von-Mises equivalent stress. The dataset and the architecture are taken from the work carried out in \cite{fernandes2025physics}.

\begin{figure}
    \centering
    \begin{subfigure}{\linewidth}
        \centering
        \includegraphics[width=\linewidth]{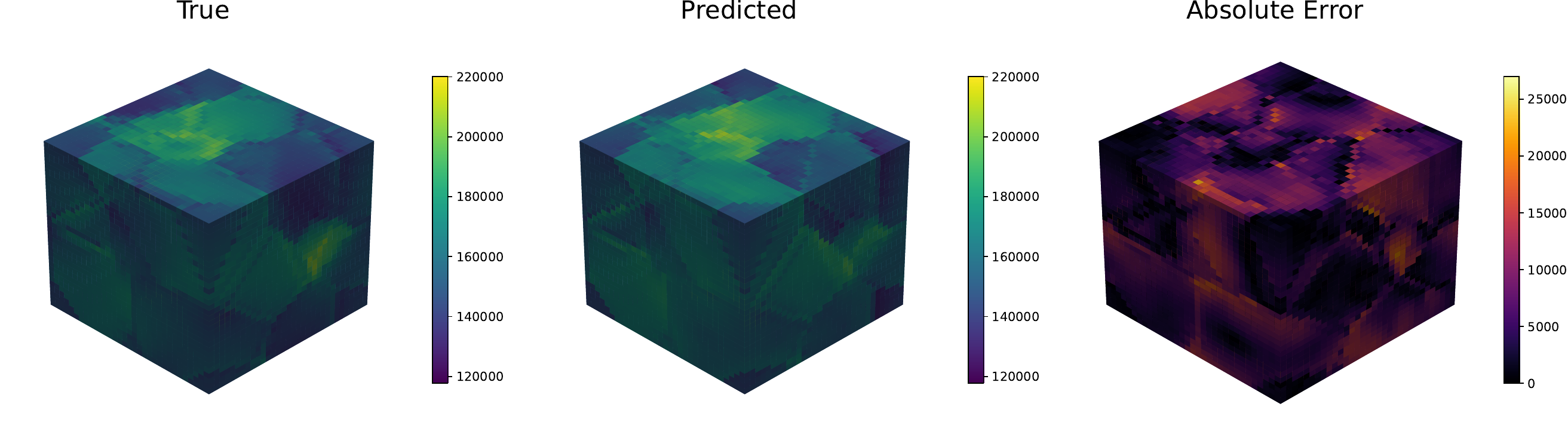}
        \caption{von-Mises equivalent stress $\sigma_{eq}$ prediction using vanilla architecture.}
    \end{subfigure}
    \begin{subfigure}{\linewidth}
        \centering
        \includegraphics[width=\linewidth]{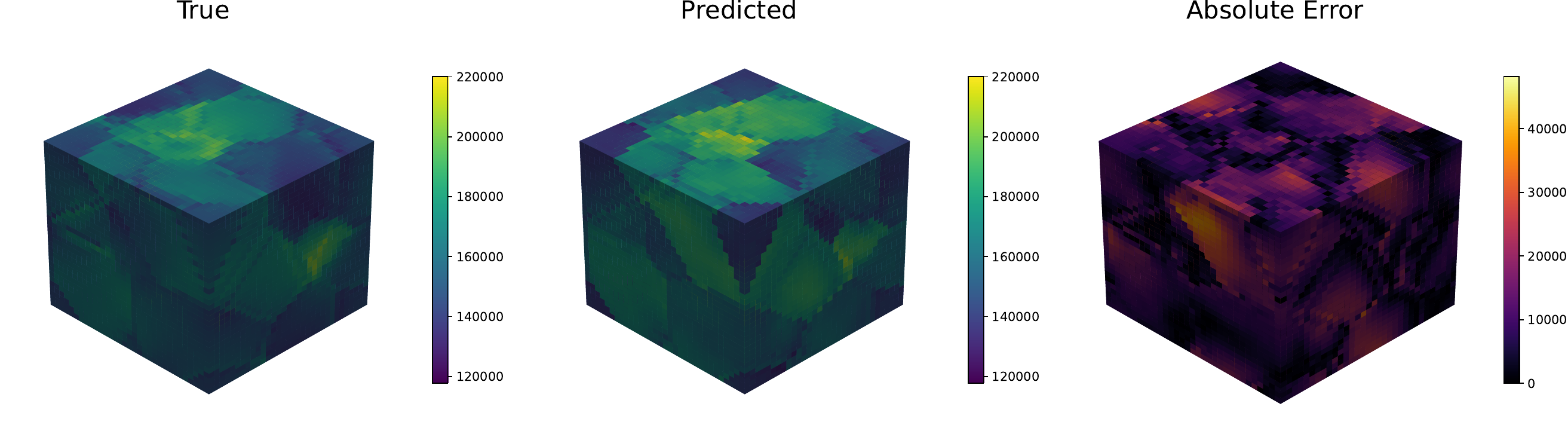}
        \caption{von-Mises equivalent stress $\sigma_{eq}$ prediction using NeuroPINN architecture.}
    \end{subfigure}
    \caption{Ground truth compared against various network predictions in E-IV.}
    \label{fig:application plots}
\end{figure}
Fig. \ref{fig:application plots} shows the ground truth compared against the predictions obtained from the vanilla architecture and the NeuroPINN architecture. As can be seen, the NeuroPINN predictions obtained give a good approximation of the ground truth. Work carried out in \cite{fernandes2025physics} reports percentage mean relative error to showcase the performance of the vanilla architecture. We follow the same metric and observe an error of $3.43\%$ when using NeuroPINN against an error of $1.68\%$ when using vanilla architecture. Barring two out of twelve spiking layers of the NeuroPINN architecture used in this example, where all neurons were observed to fire, the spiking activity observed was less than 21\%.
\section{Conclusion}\label{section: Conclusion}
Physics-Informed Neural Networks (PINNs) have emerged as a powerful alternative to traditional numerical methods for solving partial differential equations (PDEs). They leverage the expressive capacity of deep learning architectures while embedding the governing physical laws into the learning process, which often enables them to outperform purely data-driven approaches. Despite these advantages, PINNs also inherit a fundamental drawback of deep learning models: the substantial energy consumption associated with training and inference. This limitation presents a significant barrier to their deployment in resource-constrained environments, particularly in edge computing applications. To address this challenge, recent research has increasingly focused on spiking neural networks, which offer biologically inspired mechanisms for energy-efficient computation. Motivated by these developments, in this work we investigate the extension of PINN architectures into the spiking domain, with the goal of reducing computational energy requirements while preserving accuracy in solving PDEs.

To this end, we introduce Neuroscience-Inspired Physics-Informed Neural Networks (NeuroPINNs), which integrate Variable Spiking Neurons (VSNs), the Stochastic Projection method, and surrogate gradient techniques to efficiently solve the PDE under consideration. The incorporation of VSNs enables the transition of conventional PINN architectures into the spiking domain while preserving accuracy through graded spiking mechanisms. Within this framework, the SP method is employed to compute gradients for the loss function, while surrogate gradient-assisted backpropagation is utilized to optimize the trainable parameters. The key contributions and findings of this work can be summarized as follows:
\begin{itemize}
    \item NeuroPINNs paired with the SP method consistently outperform configurations that rely on surrogate backpropagation to compute the loss function. This observation reinforces the hypothesis that using surrogate gradients for loss computation may introduce inaccuracies.
    \item The performance of NeuroPINNs is comparable to that of SP-PINNs, and in Example III, NeuroPINNs even demonstrated superior accuracy relative to vanilla SP-PINNs..
    \item Employing multiple spike time steps further enhances the performance of NeuroPINNs while preserving sufficient communication sparsity, which indicates potential for energy savings.
    \item Across all four benchmark problems, NeuroPINNs outperform CPINNs. This aligns with expectations, since networks trained natively in the spiking domain tend to perform more reliably than models converted from conventional architectures.
\end{itemize}
The results demonstrate that the proposed NeuroPINNs are effective in solving PDEs of varying complexity, while operating natively in the spiking domain and enabling sparse communication. The applicability of NeuroPINNs was further demonstrated on a three-dimensional linear elastic micromechanics problem, showcasing their potential for realistic, high-dimensional scientific applications. Nonetheless, as with any research endeavor, there remains scope for further refinement of the framework. A key direction for improvement lies in replacing surrogate gradient–assisted backpropagation with a training algorithm more suitably tailored to spiking neural networks.
\section*{Acknowledgment}
SG acknowledges the financial support received from the Ministry of Education, India, in the form of the Prime Minister's Research Fellows (PMRF) scholarship. SC acknowledges the financial support received from Anusandhan National Research Foundation (ANRF) via grant no. CRG/2023/007667 and from the Ministry of Port and Shipping via letter no. ST-14011/74/MT (356529).

\end{document}